\documentclass[prd,preprint, superscriptaddress,preprintnumbers,nofootinbib]{revtex4}
\usepackage{graphicx}
\usepackage{epsfig}
\usepackage{bm}
\usepackage{latexsym,amssymb,amsmath,amssymb,wasysym,float}

\usepackage{color}

\usepackage{enumitem}
\usepackage{hyperref}

\usepackage[usenames,dvipsnames]{xcolor}
\definecolor{orange}{cmyk}{0,0.5,1,0}
\definecolor{rossoCP3}{cmyk}{0,.88,.77,.40}
\definecolor{graa}{rgb}{0.8,0.8,0.8}
\definecolor{blaa}{rgb}{0.2,0.2,0.6}

\newcommand{\beq}{\begin{equation}}
\newcommand{\eeq}{\end{equation}}
\newcommand{\be}{\begin{equation}}
\newcommand{\ee}{\end{equation}}
\newcommand{\ba}{\begin{align}}
\newcommand{\ea}{\end{align}}
\newcommand{\TR}{T_{\rm R}}
\newcommand{\TI}{T_{\rm I}}

\begin{document}

\preprint{UMN--TH--4402/24, FTPI--MINN--24/22}
\preprint{CERN-TH-2024-175}

\title{ $R^2$--Inflation Derived from 4d Strings, the Role of the Dilaton, and Turning the Swampland into a Mirage}

\author{\bf Ignatios Antoniadis}

\affiliation{High Energy Physics Research Unit, Faculty of Science, Chulalongkorn University, Bangkok 1030, Thailand}

\affiliation{Laboratoire de Physique Th\'eorique et Hautes \'Energies
  - LPTHE \\
Sorbonne Universit\'e, CNRS, 4 Place Jussieu, 75005 Paris, France
}

\author{\bf Dimitri V. Nanopoulos}

\affiliation{Academy of Athens, Division of Natural Sciences, Athens 10679, Greece
}

\affiliation{George P. and Cynthia W. Mitchell Institute for Fundamental Physics and Astronomy, Texas A\&M University, College Station, TX 77843, USA
}

\affiliation{Theoretical Physics Department, CERN, CH-1211 Geneva 23,
  Switzerland
}

\author{\bf Keith A. Olive}

\affiliation{William I. Fine Theoretical Physics Institute, School of Physics and Astronomy, University of Minnesota, Minneapolis, MN 55455, USA
}

\begin{abstract}
  \vskip 2mm \noindent 
Based on a previously derived superstring model possessing a cosmological sector that mimics Starobinsky inflation, we analyze several questions addressed in the recent literature: the generation of an effective $R^2$-term, the stability of the sgoldstino , the modular symmetry of the inflaton potential and the large distance swampland conjecture. We first show that the presence of the string dilaton stabilizes the sgoldstino direction in the supersymmetric case and no modification of the K\"ahler potential is needed. This is a generic property of a large class of Starobinsky type models within the framework of no-scale supergravity.
We then present an explicit example of a string derived inflaton potential 
where the large values of the inflaton field during inflation imply a decompactification of two extra dimensions, while the scale of inflation is generated by higher order $\alpha'$-corrections via expectation values that cancel the D-term of an anomalous $U(1)$ symmetry and break the modular symmetry of the scalar potential. As a result, the scale of inflation is much lower than the compactification scale which at the end of inflation is fixed at the free-fermionic self-dual point at an (approximate) supersymmetric minimum.

\end{abstract}

\maketitle 

\section{Introduction}

It is well known that $R + {\tilde \alpha} R^2$ gravity~\cite{Starobinsky:1980te} provides an inflationary model that lies at the heart of the observations of CMB temperature anisotropy spectrum~\cite{Planck}. It is therefore legitimate to ask if and how this simple model can be generated as an effective field theory out of a fundamental theory of quantum gravity, such as string theory. 
There are several questions that can be posed. 

\begin{itemize}
\item The first puzzle is that the theory contains a scalar degree of freedom that is not present in Einstein gravity~\cite{Stelle:1976gc}. Obviously, the string spectrum cannot change discontinuously by a quantum correction either in the inverse string tension $\alpha'$, or in the string coupling. The well known solution to this puzzle is that the scalar, often called the scalaron, can be alternatively described from a standard 2-derivative action of a minimally coupled scalar field with a very particular potential~\cite{Whitt:1984pd} 
\beq
V(\phi) =  \frac{1}{8 {{\tilde \alpha}}} 
\left(1 - e^{-\sqrt{\frac{2}{3}} \phi }
 \right)^2 \, ,
 \label{starpot}
 \eeq
where $\phi$ is the scalaron (or inflaton in the context of inflation).
The potential exponentially approaches a constant at large field values and has a minimum at $\phi = 0$ with zero vacuum energy. The inflaton mass at the minimum is $m_\phi = 1/\sqrt{6 {\tilde \alpha} }M_P$. This potential is the basis of the Starobinsky model of inflation. \footnote{Field values will always be normalized to the reduced  Planck mass with $M_P = 2.4 \times 10^{18}$~GeV.}

\item The second puzzle is that the asymptotic constant which defines the scale of inflation should be at least 5 orders of magnitude less that the Planck scale. The mass scale $m_\phi$ is fixed from the normalization of the CMB anisotropy spectrum,
\beq
A_s = \frac{V(\phi_*)}{24\pi^2 \epsilon_* M_P^4} = \frac{3 m_\phi^2}{8 \pi^2 M_P^2} \sinh^4(\phi_*/\sqrt{6}) \, , \label{As}
\eeq
where the second equality in Eq.~(\ref{As}) is specific to the Starobinsky model. 
In (\ref{As}),  $\epsilon_* = \left.\frac12 M_P^2 (dV/d\phi)^2/V^2\right|_{\phi_*}$,
and $\phi_* = 5.35$ is the value of the inflaton field at the horizon exit scale, $k_* = 0.05$~Mpc$^{-1}$ corresponding to the last 55 e-foldings of inflation, before the exponential expansion ceases at $\phi_{\rm end} \simeq 0.62$.
This leads to a tilt in CMB spectrum given by $n_s = 1 - 6\epsilon_* + 2 \eta_* = 0.9649$,
where $\eta_* = M_P^2 \left.(d^2V/d\phi^2)/V \right|_{\phi_*}$ and the tensor-to-scalar ratio, $r = 16 \epsilon_* = 0.0035$.
The Planck determined value of $A_s = 2.1 \times 10^{-9}$ implies that $m_\phi \simeq 1.25 \times 10^{-5}M_P$.

\item The third puzzle is that during inflation, the inflaton takes super-Planckian values that break the validity of the effective field theory, implying by the distance swampland conjecture the appearance of a tower of `light' states with masses exponentially small in the proper distance with an exponent of order unity in four-dimensional Planck units~\cite{Ooguri:2006in}. This tower is in general connected to the decompactification of extra dimensions or to a string tower~\cite{Lee:2019wij} implying that the `light' particles contain massive spin-2 states, which should be heavier than the Hubble constant during inflation by unitarity (Higuchi bound)~\cite{Higuchi:1986py}. This leads to an extra constraint on the inflation scale.

\item An additional puzzle is more specific to supersymmetric formulations of the Starobinsky model described by no-scale supergravity~\cite{Cecotti:1987sa, eno6,Ellis:2013nxa, kl, Ellis:2020lnc}. In these constructions of the Starobinsky potential, a second (complex) scalar field must be included in addition to the inflaton (which is of course also complex),  \cite{Cecotti:1987sa, Ellis:2013nxa}. Then, for successful inflation, the other three scalar degrees of freedom must be stabilized. If for example, 
we associate the inflaton with the real part of the `volume modulus' $T$ (as in \cite{Cecotti:1987sa}), in addition to the pseudoscalar, the second field, denoted here as $C$ must have a positive mass squared with $\langle C \rangle = 0$. However, in minimal constructions, $m_C^2 < 0$ during inflation.  
Actually $C$ is the goldstino superfield of the spontaneously broken supersymmetry during inflation.

\item As we will discuss below, a solution to the previous puzzle involves the inclusion of a third complex scalar which can be associated with the string dilaton. Then it is natural to ask, how
the inclusion of the dilaton appears in the original $R + R^2$ formulation.
\end{itemize}

In the context of string theory, an $R^2$ term cannot appear as such, since it contains an extra degree of freedom, which should be part of the free string spectrum.
Our aim therefore is to identify in the string spectrum a scalar that shares very similar properties to the Starobinsky scalaron. 
Moreover, here we restrict to $N=1$ supersymmetric compactifications which also provide a framework for interesting particle physics phenomenology incorporating a supersymmetric extension of the Standard Model. One should therefore identify within the massless string spectrum two chiral multiplets that generate a scalar potential similar to $R + R^2$ supergravity, having controllable quantum corrections, with the additional degrees of freedom (beyond the scalaron) stabilized.
 We will then attempt to provide answers to all of the above questions.

An example of such a model that was shown to address partially the first two puzzles~\cite{Antoniadis:2020txn} was constructed within the free-fermionic formulation of four-dimensional (4d) heterotic stings~\cite{Antoniadis:1986rn, Antoniadis:1987wp} whose low energy spectrum is $N=1$ supersymmetric and contains a flipped $SU(5)\times U(1)$ gauge group and three chiral families of quarks and leptons~\cite{Barr,DKN,Antoniadis:1989zy}. The gauge group breaking to the Standard Model occurs in a first order phase transition at a temperature lower than the inflation scale, implying that during inflation, the $SU(5)\times U(1)$ grand unified gauge group is unbroken~\cite{Ellis:2018moe}. The inflaton can then be identified with the superpartner of a state that mixes with the right-handed neutrinos according to the proposal of refs.\cite{Ellis:2013nxa, ENO8,Ellis:2018zya, Ellis:2020lnc}. The advantage of the free-fermionic formulation is that all string moduli are fixed at the self-dual (fermionic) point where extra symmetries arise, either local or discrete, while part of the string effective action is calculable to all orders in the inverse string tension $\alpha'$-expansion~\cite{Antoniadis:1987zk, Ferrara:1987tp, Lopez:1994ej}. Moreover, the presence of an anomalous $U(1)$, which is a general property of the Heterotic chiral models, leads to a set of vacuum expectation values (VEVs) for Standard Model singlet fields, satisfying the F- and D-flatness conditions, whose magnitudes are fixed by a natural small parameter set by the one loop anomaly~\cite{Dine:1987xk}. This defines a perturbative way to compute a new vacuum away from the initial free-fermionic one, creating calculable hierarchies in the low-energy masses and couplings~\cite{Lopez:1989fb, Rizos:1990xn, Ellis:1990vy, Kalara:1991qh, Antoniadis:1991fc, Antoniadis:2021rfm}.

The flipped $SU(5)\times U(1)$ string model was shown~\cite{Antoniadis:2020txn} to possess an inflationary sector consisting of the two necessary superfields: one contains the inflaton and the other the goldstino as its F-auxiliary component spontaneously breaks  supersymmetry, in close analogy to those linearizing $R+R^2$ supergravity~\cite{Cecotti:1987sa}. 
The corresponding two-derivative effective action was computed exactly at the string tree-level producing a scalar potential of Starobinsky type and having the same form with its supersymmetrization in $R+R^2$. 
The inflation scale is generated at 6th order in the perturbative expansion originated by the $U(1)$ anomaly mentioned above and it is naturally at the right range of energies as required by observations.

As noted above, one of the major issues in supersymmetrizing $R^2$ is that the scalar component of the goldstino superfield (sgoldstino) is unstable during inflation; its mass is tachyonic destabilizing slow-roll inflation~\cite{Ellis:2013nxa,kl}. A common approach of this problem is to modify the goldstino dependence of the K\"ahler potential leading one to abandon the nice geometric formulation of the Starobinsky model and to an arbitrariness of its supersymmetric generalization~\cite{kl, Ellis:2013nxa, Ferrara:2013wka, Antoniadis:2014oya}. One of the main results of our analysis is that this instability is absent in the presence of the string dilaton which despite its spectator role,  modifies the sgoldstino-dependence of the scalar potential and quite generally turns its mass-squared positive at the global minimum and during inflation.

Another issue specific to large-field inflation, such as in the Starobinsky model, is the breakdown of validity of the effective field theory and the appearance of a tower of light states according to the Swampland distance conjecture. To be concrete, we test this conjecture in the flipped $SU(5)\times U(1)$ string model, whose two-derivative effective action in the inflation sector is exact to all orders in $\alpha'$, and analyze its consequences to inflation. We find that the tower of light states corresponds to a Kaluza-Klein (KK) tower of two internal dimensions with a compactification scale around two orders of magnitude below the string scale, which in this construction is of the same order as the species scale (or the six-dimensional gravity scale). This is much higher than the inflation scale which is generated at higher order in perturbation theory away from the free-fermionic point, as described above.
Finally, the inflaton is associated with an $SL(2,Z)$ modular symmetry which is spontaneously broken by the VEVs which cancel the $U(1)$ anomaly and thus the scalar potential is not modular invariant. Despite this fact, we find a similar prediction as in other frameworks that relates the number of e-folds during inflation with the number of species~\cite{Casas:2024jbw}. 

Based on the above example, in this work we analyze several questions addressed in the recent literature, such as the generation of an effective $R^2$-term, the stabilization of the additional scalars,  the large distance swampland conjecture~\cite{Scalisi:2018eaz, Rasulian:2021wny, Lust:2023zql} and the modular symmetry of the inflaton potential~\cite{Casas:2024jbw, Kallosh:2024ymt}, answering all of the puzzles we described above. In what follows, we first briefly review the construction of the Starobinsky potential from $R + R^2$ gravity (with and without a dilaton), its supersymmetrization, and briefly review an explicit model \cite{Antoniadis:2020txn} in Section \ref{sec:2}. Then, in Section \ref{sec:3}, we show how the inclusion of the dilaton stabilizes the sgoldstino at the global minimum. We generalize this mechanism to a wider class of no-scale models in Section \ref{sec:4} and discuss the conditions where stabilization is maintained away from the minimum, during inflation. 
We discuss the implications of these models for the large distance conjecture in Section \ref{sec:5}.
We summarize our results in Section \ref{sec:6}.

\section{Supersymmetrization of $R^2$ and the string dilaton}
\label{sec:2}
\subsection{The Starobinsky model}

For pedagogical reasons, we first briefly review the classical equivalence between the Starobinsky model $R+{\tilde \alpha} R^2$ and a scalar-tensor theory with a particular scalar potential. Therefore we consider the action
\begin{equation}
 S \; = \;  \int d^4x\, \sqrt{-{\tilde g}} \left( \frac{1}{2} R[\tilde{g}] + \frac{{\tilde \alpha}}{2} R[\tilde{g}]^2 \right) \, ,
\label{EH}
\end{equation}
where $\tilde{g}$ is the `Jordan' frame metric (to be clarified below).
To transform the action to the Einstein frame, we first introduce a Lagrange multiplier $\Phi$ identifying the scalar curvature $\tilde{R}$ with a scalar field $\chi$:
\be
\label{starbos1}
S=\int[d^4x]\left\{{1\over 2}\tilde{R}+\Phi(\tilde{R}-\chi)+\frac{{\tilde \alpha}}{2}\chi^2\right\} \,,
\ee
where the brackets in the integration measure include the density factor $\sqrt{|\det \tilde{g}|}$ and in our conventions the signature of the metric is mostly positive. Again, we remind the reader that we work in reduced Planck mass units.
We proceed by integrating over $\chi$ which has a Gaussian dependence:
\be
\label{starbos2}
S=\int[d^4x]\left\{{1\over 2}(1+2\Phi)\tilde{R}-{1\over 2{\tilde \alpha}}\Phi^2\right\}
\ee
and perform a Weyl rescaling of the metric 
\begin{equation}
g_{\mu \nu}  \; = \; e^{2\Omega} \tilde{g}_{\mu \nu} \; = \;  \left(1 + 2 \Phi \right) \tilde{g}_{\mu \nu} \, ,
\label{weyl}
\end{equation}
to find:
\begin{equation}
S \; = \;   \int [d^4x]  \left[ \frac{1}{2} R[g] - \frac{3} {(1 + 2   \Phi)^2} \partial^\mu \Phi \partial_\mu \Phi -  \frac{\Phi^2}{2 {{\tilde \alpha}}(1 + 2   \Phi)^2}  \right] \, .
\label{almostStaro}
\end{equation}
After a field redefinition, $\phi = \sqrt{3/2}\ln (1+2\Phi)$ we arrive at
the standard form of the Starobinsky scalar potential:
\be
\label{starbos4}
S=\int[d^4x]\left\{{1\over 2}R-{1\over 2}(\partial\phi)^2-{1\over 8{\tilde \alpha}}\left(1-e^{-{\sqrt{2/3}\,\phi}}\right)^2\right\}\,,
\ee
where the indices of partial derivatives are contracted with the metric in shorthand notation. This action now contains the Starobinsky potential in Eq.~(\ref{starpot}).

Note that in all steps above from \eqref{starbos1} to \eqref{starbos4}, no constraint on the parameter ${\tilde \alpha}$ was used and the above result is valid also if ${\tilde \alpha}$ is field dependent.

Indeed, this observation is very important in order to introduce the string dilaton whose VEV determines the string coupling $g_s$. Since the string spectrum is determined at tree-level, the dilaton dependence in the string frame should be a common factor of the string effective action acting as $1/g_s^2$:
\be
\label{stringS1}
S_\text{string frame}=\int[d^4x]e^{-2\varphi}\left\{{1\over 2}R[G]+2(\partial\varphi)^2+{\cal L}_\text{matter}\right\}\,,
\ee
where $G$ is the string frame ($\sigma$-model) metric, $\varphi$ is the dilaton with $\langle e^\varphi\rangle = g_s$, while ${\cal L}_\text{matter}$ denotes a matter Lagrangian independent of $\varphi$, as are the last two terms of \eqref{starbos4}. One can then go to the Einstein frame by a Weyl rescaling of the metric $G=e^{2\varphi}g$, leading to:
\be
\label{stringS2}
S_\text{string}=\int[d^4x]\left\{{1\over 2}R[g]-(\partial\varphi)^2-{1\over 2}(\partial\phi)^2
-{e^{2\varphi}\over 8{\tilde \alpha}}\left(1-e^{-{\sqrt{2/3}\,\phi}}\right)^2\right\}\,,
\ee
where 
the parameter ${\tilde \alpha}$ can now be treated as a numerical constant.
Reversing the manipulations we have used to pass from \eqref{starbos1} to \eqref{starbos4}, one rewrite \eqref{stringS2} in a geometric form:
\be
\label{stringS3}
S_\text{string}=\int[d^4x]\left\{{1\over 2}R[\tilde{g}]-(\tilde{\partial}\varphi)^2 +\frac{{\tilde \alpha}}{2} e^{-2\varphi}\left[\tilde{R}-2(\tilde{\partial}\varphi)^2\right]^2 \right\}\,.
\ee
The interested reader can verify \eqref{stringS2} starting from \eqref{stringS3} by following exactly the steps \eqref{starbos1} to \eqref{starbos4} and replacing $\tilde{R}$ by $\left[\tilde{R}-2(\tilde{\partial}\varphi)^2\right]$ in \eqref{starbos1}, \eqref{starbos2} and $R$ by $\left[R-2({\partial}\varphi)^2\right]$ in \eqref{almostStaro}, \eqref{starbos4} recalling that both $R$ and $(\partial \varphi)^2$ transform with the same conformal factor $e^{2\Omega}$ which is still given by Eq.~(\ref{weyl}).

Note that the dilaton exponent in \eqref{stringS3} is consistent with a tree-level $R^2$ which is (globally) scale invariant and does not change under metric rescalings (modulo four-derivative interactions of the dilaton and metric).

\subsection{Supersymmetrization}

The supersymmetrization of $R+{\tilde \alpha} R^2$ has been performed in \cite{Cecotti:1987sa, Ferrara:2013wka, Antoniadis:2014oya}. One might naively expect to be able to describe it in the context of ordinary $N=1$ supergravity coupled to a chiral multiplet corresponding to a super-Lagrange multiplier, following a procedure similar to the one described above in the bosonic theory. It turns out however, that one needs to introduce two Lagrange multipliers chiral superfields.\footnote{This is related to comment made earlier regarding the construction of the Starobinsky potential in no scale supergravity requiring (at least) two chiral superfields \cite{Ellis:2013nxa}.} The reason is the the scalar curvature appears in the upper component of a chiral superfield $\cal R$~\cite{wesbag} and therefore ${\cal R}^2$ does not contain $R^2$. The latter appears in a D-term of ${\cal R}\bar{\cal R}$ whose linearization requires two super-Lagrange multipliers.
The result is an $N=1$ supergravity theory coupled to two chiral multiplets, $T$, and $C$ with a K\"ahler potential of the no-scale type~\cite{Cremmer:1983bf, Ellis:1983ei,Lahanas:1986uc} and superpotential given by:
\be
\label{susy1}
K=-3\ln(T+{\bar T}-C{\bar C})\,,\quad W=M C\left(T-{1\over 2}\right)
\ee
where $M^2 = 3/(2{\tilde \alpha})$~\footnote{Parallels between the $R+{\tilde \alpha} R^2$ theory and no-scale supergravity were discussed in \cite{DLT,DGKLP,eno9,Ellis:2020lnc,Ema:2024sit}}. The scalar component of the superfield $T$ corresponds to the Starobinsky inflaton with  $\TR=\text{Re}T=\frac12 e^{\sqrt{2/3}\phi}$, with the same scalar potential \eqref{starpot} when $\TI=\text{Im}T=C=0$.
The superfield $C$ contains the goldstino during inflation, as can be seen from the linear term in the superpotential \eqref{susy1}. 

In fact, the full scalar potential is given by:
\be
\label{susy2}
V = \frac{M^2}{12 (T + {\bar T} - C {\bar C})^2} \left\{ 1 - 2 (T + {\bar T}) + 4 T  {\bar T}  + [ 8-  4 (T + {\bar T}) ] C  {\bar C} \right\} \, .
\ee
With the above relation between $\TR$ and $\phi$ with $C=0$, one immediately recovers the Starobinsky potential. 
 The potential has a global supersymmetric minimum at zero energy at (the self-dual point) $\TR=1/2$ ($\phi = 0$), $\TI=C=0$. The mass of the inflaton at the minimum is given by $m_\phi=M/3$.
As discussed above, for this model to reproduce CMB data, we require
$M = 3.75 \times 10^{-5} M_P$ and as noted earlier, $\phi_* = 5.35$ corresponding to $\TR \simeq 40$
for $N_* = 55$.

Note that $C$ is tachyonic during inflation creating an instability. Indeed the mass-squared of $C$ and $\TI$ read:
\begin{align}
\label{susy3}
m_C^2 &={T+{\bar T}\over 3}{\partial^2 V\over{\partial C\partial{\bar C}}}{\bigg|}_{C={\bar C}=0}
= M^2\, {1+2(T+{\bar T})-2(T^2 + {\bar T}^2)\over 18\, (T+{\bar T})^2}\nonumber \\
m_{\text{Im}T}^2 &= {(T + {\bar T})^2\over 3} {\partial^2 V\over ({\partial\text{Im}T})^2} ={2\over 9}M^2
\,,
\end{align}
where the factors $(T+{\bar T})/3$ and $(T + {\bar T})^2/3$ come from the normalization of the kinetic terms. It follows that $\TI$ has a large positive mass-squared and we can safely set $\TI = 0$. In this case, $C$ becomes tachyonic for $\TR>(\sqrt{2}+1)/2 \simeq 1.2$.
This corresponds to $\phi \simeq 1.1< \phi_*$, making the inflationary trajectory unstable to any fluctuations in the $C$ (or ${\bar C}$) direction.
One way to cure this instability during inflation is to modify the K\"ahler potential by adding a quartic $(C{\bar C})^2$ in the argument of the logarithm \cite{Ellis:1984bs,Ellis:2013nxa,Ellis:2015kqa,Ellis:2018zya,Kaneta:2019yjn,Ellis:2020lnc,Ellis:2020xmk}. But this comes at the expense of losing the geometric interpretation of the (supersymmetric) Starobinsky model. Another approach is to parametrize the model dependent stabilization of the sgoldstino by integrating it out to an effective field theory obtained by imposing a non-linear supersymmetric nilpotent constraint $C^2=0$ which eliminates the scalar component in terms of the goldstino bilinear~\cite{Antoniadis:2014oya}. Here, we choose to not follow this approach since we would like to keep $N=1$ linear supersymmetry at low energies, as we described in the introduction. Instead, we discuss an alternative solution to the stabilization of $C$ in Section~\ref{sec:3}.

\subsection{An explicit string model}

Here we present a short review of the string model~\cite{Antoniadis:2020txn}, constructed within the free-fermionic formulation of 4d heterotic superstrings, that shares similar properties (although not identical) with the $R^2$ supergravity model described above. In particular, it contains the dilaton which is perturbatively undetermined, although all other moduli are fixed at the fermionic self-dual point where extra symmetries (gauge and discrete) appear. For making the comparison transparent, it is therefore convenient to change the field variables $T$ and $C$ to an appropriate `charged' or `symmetric' basis $y$ and $z$: 
\be
\label{charge1}
T={1\over 2} \left({1+y\over 1-y} \right)\quad;\quad C={z\over 1-y}\,,
\ee
so that \eqref{susy1} becomes:
\be
\label{charge2}
K=-3\ln\left(1-|y|^2-|z|^2\right)\quad;\quad W=Mzy( 1-y)
\ee
and the inflationary region of large $\TR$ is mapped to the boundary of the K\"ahler domain $|y|\to 1$ while $C=0$ corresponds to $z=0$. The scalar potential \eqref{susy2} in these coordinates becomes:
\be
\label{charge3}
V=\frac{M^2}{3}
{|y|^2 |1-y|^2 + |z|^2 (1-2(y+{\bar y})+3|y|^2)\over (1-|y|^2-|z|^2)^2}
\ee
Along the direction $y_I= z=0$ we can redefine $y_R = \tanh(\phi/\sqrt{6})$, where $y\equiv y_R+iy_I$ and $\phi$ is the canonically normalized inflaton. We then obtain the same potential as in Eq.~(\ref{starpot}) with $M^2=3/(2{\tilde \alpha})$ (as it must, since we simply performed a field redefinition). Now, $\phi_* = 5.35$, corresponds to $y_R \simeq 0.975$.

As noted, successful inflation requires $y_I= z=0$, and therefore these directions must be stable. Although the direction $y_I=0$ is stable, $z$ is tachyonic during inflation, and must be stabilized as was the case for $C$ in the former basis.

The K\"ahler potential of the string model is~\cite{Antoniadis:2020txn}:
\be
\label{charge4}
K_\text{string}=-\ln(S+{\bar S})-2\ln\left(1-|y|^2\right)-2\ln\left(1-{1\over 2}|z|^2\right)\,,
\ee
which is similar, but somewhat different from \eqref{charge2} and was shown to be exact at the string tree-level, to all orders in $\alpha'$. The superpotential is unchanged from \eqref{charge2} and the inflation mass scale $M$ is generated at 6th order of non renormalizable terms, via non-trivial VEVs of $SU(5)\times U(1)$ singlets driven by the anomalous $U(1)$. If we ignore for now the contribution from the dilaton, $S$, the resulting scalar potential, although not identical, shares the same properties as the potential \eqref{starpot}. In this case it is given by
\be
\label{VANR}
V=M^2
\frac{4|y|^2 |1-y|^2 + 2|z|^2 (1-2(y+{\bar y})+2|y|^2+2(y+{\bar y})|y|^2-3|y|^4)+|y|^2 |1-y|^2|z|^4}{(1-|y|^2)^2(2-|z|^2)^2}\, .
\ee
For $y_I= z=0$, this potential reduces dramatically to 
\beq
V = M^2 \frac{y_R^2}{(1+y_R)^2}\, .
\eeq
Then redefining $y_R = \tanh(\phi/2)$, we obtain
a Starobinsky-like potential
\beq
V = \frac{M^2}{4}(1-e^{-\phi})^2 \, .
\label{ANRpot}
\eeq
The mass of the inflaton at the minimum $\phi=0$ is in this case $m_\phi = M/\sqrt{2}$.
Moreover, the amplitude of primordial density perturbations during inflation is similar to Eq.~(\ref{As})
\beq
A_s = \frac{M^2}{12 \pi^2} \sinh^4(\phi/2) \, ,
\eeq
which gives $M = 1.75 \times 10^{-5} M_P$. The pivot scale for this model is $\phi_* = 4.75$ which then determines $n_s = 0.9645$,
very close to the value in the Starobinsky model,
and $r = 0.0024$ which is slightly lower than that in the Starobinsky model. 

The potential (\ref{ANRpot}) is actually part of a larger class of models known as $\alpha$-Starobinsky or no-scale attractor models \cite{Ellis:2013nxa,KLR,Kallosh:2014rga,Roest:2015qya,Scalisi:2015qga,ENOV3,Ellis:2020xmk,Saini:2024mun}.
In the context of no-scale supergravity, we can replace the coefficient of the logarithm in Eqs.~(\ref{susy1}) and (\ref{charge2}), $3 \to 3\alpha$. 
Indeed, it was shown in \cite{Ellis:2013nxa}, that for a potential $\propto (1-e^{-\sqrt{2/3\alpha} \phi})^2$ leads to a value of $n_s$ independent of $\alpha$, whereas $r \propto \alpha$. 
It is clear then that the $R+{\tilde \alpha} R^2$ construction requires $\alpha = 1$, whereas the model described by Eq.~(\ref{charge4})
corresponds to $\alpha = 2/3$ and accounts for the difference in the predicted value of $r$.

However, as in the previous example, we see that $z$ is tachyonic during inflation.
The mass squared of $z$ is
\beq
m_z^2 ={\partial^2 V\over{\partial z\partial{\bar z}}}{\bigg|}_{z={\bar z}=0}
= M^2\, \frac{1-2(y+{\bar y})+4|y|^2 -|y|^4} { 2(1-|y|^2)^2} \, .
\eeq
Along the $y_I=0$ direction, 
it is easy to see, that $z$ is tachyonic for $y>\sqrt{2}-1$, i.e. during inflation. 
As we will see in the next section, this problem can be resolved when we include the $\ln(S+{\bar S})$ term in Eq.~(\ref{charge4}).
While the mass-squared of $y_I$ is not constant during inflation, $y_I = 0$ is always a minimum.

\section{Stabilization and the Role of the Dilaton}
\label{sec:3}

In this section, we will first 
see that  the inclusion of the dilaton  can stabilize the $C$ (or $z$) degrees of freedom in the supersymmetric version of $R+R^2$ for all inflaton field values including during inflation. We will then show 
that the inclusion of $S$  can also stabilize the $z$ degree of freedom in the model described by Eq.~(\ref{charge4}).
This solution requires of course that the dilaton itself is stabilized 
which goes beyond the scope of this paper though we  comment on this before concluding. 

The presence of the string dilaton can be incorporated in a straightforward way, following the steps \eqref{stringS1}-\eqref{stringS3} within the bosonic theory above. More precisely, the string dilaton is part of a chiral superfield $S$ with $S_R=e^{-2\varphi}$, while $S_I$ is the universal axion, dual to a 2-index gauge potential. It follows that the dilaton dependence amounts a modification of the K\"ahler potential \eqref{susy1} by an extra term $K\to K+\delta K$ given by:
\be
\label{dil1}
\delta K=-\ln(S+{\bar S})\,.
\ee
The resulting scalar potential (using the superpotential in Eq.~\eqref{susy1}) is modified as:
\begin{align}
\label{dil2}
V = & \frac{M^2}{12 (S+{\bar S} ) (T + {\bar T} - C {\bar C})^3} \times \nonumber \\ 
& \left\{ (T + {\bar T})|2T-1|^2 + 2 C{\bar C} \left[1+ 2(T + {\bar T}) -2 (T^2 + {\bar T}^2) \right] + 4 (C  {\bar C})^2 (T + {\bar T} -2) \right\}
\end{align}
Along the direction $\TI = C = {\bar C} = 0$, the potential reduces to 
\beq
V = \frac{M^2}{48 (S+{\bar S})} \frac{(1-2\TR)^2 }{\TR^2}\, ,
\eeq
which after the transformation 
$\TR = \frac12 e^{\sqrt{2/3}\phi}$ is the potential in Eq.~(\ref{starpot})
but now divided by $(S+{\bar S})$.

It can now be easily seen that the tachyonic direction of the sgoldstino $C$ is lifted and no instability is present during inflation. Indeed the $C$ and $\TI$ masses now become:
\begin{align}
\label{dil3}
m_C^2{\bigg|}_{C=0} &= M^2\, {5-2(T+{\bar T})-4(T^2+{\bar T}^2)+12T {\bar T}\over 36(S+{\bar S})(T+{\bar T})^2}\nonumber \\
m_{\TI}^2 &= {2\over 9(S+{\bar S})}M^2\,,
\end{align}
so that for $\TI=0$, $C$ is not tachyonic:
\be
\label{dil4}
m_C^2\bigg|_{\TI=0}={M^2\over 36(S+{\bar S})\TR^2}\left[\left(\TR-{1\over 2}\right)^2+1\right]>0\,.
\ee
Therefore, the presence of the dilaton can cure the instability in the sgoldstino direction and no modification of the K\"ahler potential is needed. During inflation, the inflaton  mass is
\be
\label{inflatonmass}
m_\phi^2{\bigg|}_{C=\TI=0} \simeq -{M^2\over 9(S+{\bar S})} e^{-\sqrt{2\over 3}\phi_*}
\ee
in the limit $\phi_*\gg1$.
It follows that $C$ and $\TI$ are much heavier and can be set to zero, ie. placed at their minima. 
The potential still has a supersymmetric global minimum at zero energy at $\TR=1/2$, $\TI=C=0$ where $S$ is a flat direction.
Of course, the dilaton has to be stabilized fixing the string coupling otherwise the scalar potential during inflation is runs away in the dilaton direction. 

For completeness, we provide the potential in the $y,z$ variables
\be
\label{chargeS3}
V={M^2\over 3(S+{\bar S})}
{|y|^2 |1-y|^2 (1-|y|^2) + |z|^2 (1-2(y+{\bar y})+4|y|^2-|y|^4-|z|^2(1-2(y+{\bar y})+3|y|^2))\over (1-|y|^2-|z|^2)^3} \, ,
\ee
which not surprisingly gives us again the Starobinsky potential (modulo the factor of $(S+{\bar S})^{-1}$) when $y_R = \tanh(\phi/\sqrt{6})$.

Let us now return to the string model with the K\"ahler potential given in Eq.~(\ref{charge4}) including the dilaton.
\be
\label{VANR}
V=M^2
\frac{4|y|^2 |1-y|^2 + 2|z|^2 (1-2(y+{\bar y})+4|y|^2-|y|^4)+|y|^2 |1-y|^2|z|^4}{(S+{\bar S})(1-|y|^2)^2(2-|z|^2)^2}\, .
\ee
which reduces to Eq.~(\ref{ANRpot}) using $y_R = \tanh(\phi/2)$ along $z = {\bar z} = y_I = 0$ with the multiplicative factor $(S + {\bar S})^{-1}$. 

As in the case of the Starobinsky model, we see that the mass-squared of $z$ and $y_I$ are
\begin{eqnarray}
\label{dil5}
m_z^2{\bigg|}_{z=0} &=& M^2\, {1-2(y+{\bar y})-2|y|^2 (y+{\bar y})+6|y|^2 + |y|^4 \over 2(S+{\bar S})(1-|y|^2)^2}\\
m_{y_I}^2{\bigg|}_{z=0} &=& M^2{(1-y_R+2y_R^2)\over 1+y_R}>0\,. 
\end{eqnarray}
It is easy to see that for $y_I = 0$ 
\be
\label{dil6}
m_z^2\bigg|_{y_I=0}={M^2\over 2(S+{\bar S})} \frac{(1-y_R)^2}{(1+y_R)^2}>0
\ee
is clearly positive definite thus stabilizing the $z$-direction all along the inflationary trajectory. Moreover during inflation, the inflaton mass is given by (see \eqref{ANRpot})
\be
\label{inflatonmassstring}
m_\phi^2{\bigg|}_{C=\text{Im}T=0} \simeq -{M^2\over 2(S+{\bar S})} e^{-\phi}
\ee
and as before $z$ and $y_I$ are much heavier and can be set to zero, as $y_I=0$ is always minimum of the potential (for $z = 0$). Thus, the need for quartic corrections to the K\"ahler potential is removed. 
In the next section, we will consider some general conditions for which the inclusion of the dilaton will lead to the stabilization of the sgoldstino.

\section{Generalization in a wider class of no-scale supergravity models}
\label{sec:4}

In the previous section, we saw that the simple inclusion of the dilaton in the K\"ahler potential provided a positive mass-squared contribution to the sgoldstino, removing its tachyonic behavior. This was shown explicitly for the Starobinsky and string-derived model. 
In this section, we show how this occurs in wider class of models.

Let us consider first the K\"ahler potential given in Eq.~(\ref{susy1}). Let us further take a superpotential of the form
\beq
W = M C f(T) \, .
\label{phif}
\eeq
The scalar potential is
\beq
V = \frac{M^2}{3(T+{\bar T} - C {\bar C})^2}  \left[ |f(T)|^2 + |C|^2 \left( (T+{\bar T}) |f_T|^2 - 2 (f {\bar f}_{\bar T} + {\bar f} f_T) \right) \right] \, .
\eeq
As before, $\TR$ will be associated with the inflaton. Then along the direction $\TI = C= 0$,
the scalar potential is simply
\beq
V = \frac{M^2}{12\TR^2} f(T)^2 \, .
\label{Vsimp}
\eeq
With this ans\"atz, both $\TI = 0$ and $\phi = 0$ are extrema. For $\TI = 0$, to be a minimum, $f_T^2 - f f_{TT}$ must be positive ($f_T = df/dT$). This is of course true for the Starobinsky model derived from Eq.~(\ref{susy1}) with $f(T) = T-1/2$. 
For the sgoldstino, the direction $C=0$ is also an extremum, and the condition for a minimum (along $\TI = 0$) is $f^2 - 4T f f_T + 2 T^2 f_T^2> 0$.
However, during inflation or wherever the potential is nearly flat, $T f_T \simeq f$ in which case 
\beq
m_C^2 = -\frac{M^2}{6 \TR^3} f(T)^2 \qquad T = \TR  \, ,
\label{Cf}
\eeq
that is during inflation, $C$ is necessarily tachyonic, requiring some form of stabilization if inflation is to occur.

Note that this model is equivalent to a `dual' model obtained by replacing in the superpotential \eqref{phif}, $f(T)$ with ${\tilde f}(T)=(\lambda T)^2f({1\over\lambda T})$, with $\lambda$ an arbitrary constant. This can be easily seen using the fact that $T\to 1/(\lambda^2 T)$ and $C\to C/(\lambda T)$ amounts to a K\"ahler transformation $K\to K+3\ln(\lambda T)+3\ln(\lambda{\bar T})$. In the case of $f(T)=T-1/2$, choosing $\lambda=2$, one finds that ${\tilde f}(T)=2T(T-1/2)$ and the canonical field is $\TR = \frac12 e^{-\sqrt{2/3}\phi}$.

A similar property holds for $\alpha$-attractors with K\"ahler potential
\beq
K = -3\alpha \ln (T+{\bar T} - C {\bar C})\, .
\eeq
A superpotential of the form  \cite{Kallosh:2014rga}
\beq
W = \sqrt{\alpha} M C f(T) (2 T)^\frac{(3\alpha-3)}{2} \, ,
\eeq
will reduce to Eq.~(\ref{Vsimp})
along $\TI = C = 0$. 
The relation between $T$ and the canonical $\phi$ is altered to $T = \frac12 e^{\sqrt{2/3\alpha}\phi}$.
For the choice $f = T-1/2$,  the potential becomes 
\beq
V = \frac{M^2}{12} \left( 1 -e^{-\sqrt{2/3\alpha}\phi}\right)^2 \, .
\eeq
The condition for $\TI$ to be a minimum is now altered. It is  $f_T^2 - f f_{TT}  + \frac32(\alpha - 1) f^2/T^2 > 0$.
For the ($\alpha$)-Starobinsky choice $f = T-1/2$, $\TI = 0$ is always a minimum if
$\alpha \ge 1$. For $1/3 < \alpha < 1$, there is a tachyonic instability appearing at small $\TR < 1/4$ corresponding to 
$\phi_{\rm tach} < -\ln 2/\sqrt{2}$. However, this instability is actually never reached as during its oscillations $|\phi_{\rm tach}| > \phi_{\rm end}$,
where the latter is the inflaton field value when exponential expansion ends and oscillations begin. For $\alpha = 1$, $\phi_{\rm end} \simeq 0.62$,
whereas for $\alpha = 1/3$, $\phi_{\rm end} \simeq 0.49$ and the amplitude of oscillations decreases with each oscillation as
$|\phi|^2 \propto a^{-3}$, where $a$ is the cosmological scale factor. For $\alpha < 1/3$, there is another tachyonic instability which occurs at
large $T$ and hence may occur during inflation.  The condition for $C=0$ to be a minimum is unchanged from the previous set of models,
and during inflation, the mass-squared of $C$ is still given by (\ref{Cf}). \footnote{For the dual model, $\TI = 0$ is always a minimum so long as $\alpha \ge 1/3$. For smaller $\alpha <1/3$, there is again a tachyonic instability, which now occurs at small $\TR$, which in this case is during inflation. }

The inclusion of the dilaton can stabilize the $C = 0$ direction in both of the above generalizations.  Starting again
with the superpotential (\ref{phif}),  and K\"ahler potential
\beq
\label{Kdil}
K=-3\ln(T+{\bar T}-C{\bar C}) - \ln(S+{\bar S})
\eeq
the scalar potential is modified and becomes 
\begin{align}
V = & \frac{M^2}{3(S+{\bar S})(T+{\bar T}-C{\bar C}))^3}  \left[ (T+{\bar T}) |f(T)|^2 + 2|C|^2 \left( |f(T)|^2 - \right.\right. \nonumber \\
& \qquad  \left. \left. (T+{\bar T}-|C|^2)  (f {\bar f}_{\bar T} + {\bar f} f_T)  +\frac12 \left( (T+{\bar T})^2-|C|^2 (T+{\bar T})\right) f_T   {\bar f}_{\bar T} \right)    \right] \, ,
\end{align}
which reduces to (\ref{Vsimp}) modulo the factor of $(S+{\bar S})^{-1}$ for $C = 0$.
Although the extremal condition for $T_I = 0$ is unchanged, the condition for a minimum at $C=0$ is now
 $5 f^2 - 8 T f f_T + 4 T^2 f_T^2 > 0$. But now during inflation, using $T f_T \simeq f$ ,
 instead of (\ref{Cf}), we have
 \beq
m_C^2 = \frac{M^2}{24 S_{\rm R}  \TR^3} f(T)^2 \qquad T = \TR  \, ,
\label{CfS}
\eeq
which is positive definite. 

Similarly for the $\alpha$-Starobinsky models, the dilaton does not affect the conditions for the minimization 
along $\TI = 0$. However, now the condition for minimization along $C=0$ is $(2+3\alpha)f^2 - 8 T f f_T+4 T^2 f_T^2$.
During inflation, the mass-squared of $C$ becomes 
\beq
m_C^2 = \frac{M^2}{24 S_{\rm R}  \TR^3} (3\alpha-2) f(T)^2 \qquad T = \TR  \, .
\label{CfaS}
\eeq
Thus dilaton stabilization of $C$ requires $\alpha > \frac23$.

\section{The scale of inflation, large field excursions and the tower of light states}
\label{sec:5}

It is known that Starobinsky inflation is of large field type implying a breakdown of the effective field theory at a scale below the Planck mass. In particular, as we mentioned earlier, the inflaton value at horizon exit $\phi_*$ is about 5 in Planck units, 
implying the existence of a tower of light states according to the swampland distance conjecture~\cite{Ooguri:2006in}. 
In order to identify the origin and nature of the tower, it is sufficient to focus on the K\"ahler potential and its emergence within the string effective action, since the superpotential is suppressed by the inflation scale proportional to $M$, which is fixed by CMB observations using Eq.~(\ref{As}) with $m_\phi \propto M$ to be about 5 orders of magnitude below the Planck scale. By inspection of its form \eqref{charge4}, inflation takes place near the boundary of the K\"ahler domain $|y|\to 1$, equivalent to large $\TR$. 

Actually, the inverse transformation \eqref{charge1} 
\be
\label{ytoT}
y={2T-1\over 2T+1}\,,
\ee
applied to \eqref{charge4} and to the superpotential \eqref{charge2} yields:
\be
\label{TzBasis}
K_\text{string}=-\ln(S+{\bar S})-2\ln(T+{\bar T})-2\ln\left(1-{1\over 2}|z|^2\right)\quad;\quad W=M z\left(T-{1\over 2}\right)\,.
\ee

We recall that the $y$- and $z$-dependent part of the K\"ahler potential corresponds to fields associated  
with one of the three planes of $T^6$ which takes the factorized form $T^2\times T^2\times T^2$, where all fields are at the free fermionic point. Following the derivation of the $N=1$ effective supergravity of four-dimensional superstrings constructed within the free fermionic formulation and translated in a basis of a $\mathbb{Z}_2\times\mathbb{Z}_2$ orbifold~\cite{Antoniadis:1987zk, Ferrara:1987tp}, $z$ is an untwisted field from the third plane while $y$ is twisted in the first two planes~\cite{Antoniadis:2020txn}. Because of the symmetry between all fields defined around the fermionic point, the boundary of the K\"ahler domain corresponds by the field redefinition $y\to T$  in \eqref{ytoT} to a decompactification limit where the area of the third $T^2$ $(\TR)$ becomes large. Indeed $2T=\sqrt{G}+ib$ in string units, where $G$ is the determinant of the two-dimensional metric of the third plane and $b$ is the corresponding 2-index antisymmetric tensor which in 2 dimensions has one element. At the fermionic point, the complex structure is unity, corresponding to a square torus of radius $R$, and thus $\sqrt{G}=R^2$.

From the K\"ahler potential \eqref{TzBasis}, one finds the kinetic term for $R$:
\be
\label{kinTstring}
-2{\partial T\partial\bar{T}\over (T+\bar{T})^2}=-2\left({\partial R\over R}\right)^2-{1\over 2}{(\partial b)^2\over R^2}\,.
\ee
This can be compared to the action one obtains for $R$ via dimensional reduction of the Einstein-Hilbert action in $4+d$ dimensions compactified on a $d$-dimensional square torus of radius $R$:
\be
\label{dimredd}
{1\over 2}{\cal R}^{(4+d)} \longrightarrow{1\over 2}{\cal R}^{(4)}-{d(d+2)\over 4}\left({\partial R\over R}\right)^2\,.
\ee
Agreement of \eqref{dimredd} with \eqref{kinTstring} implies $d=2$, consistent with the string theory argument above.
The canonically normalized inflaton field is $\phi=2\ln R$ and takes a value around 5 in Planck units during inflation, as mentioned above. It follows that $RM_*\sim e^{2.5}$ (corresponding to $T_R\simeq 75$) where $M_*$ is the string scale satisfying:
\be
\label{M*}
M_p^2={1\over g_s^2}M_*^2(RM_*)^d\,.
\ee
Thus $RM_*=(g_sM_p/M_*)^{2/d}$, implying
\be
\label{RMp}
RM_p={1\over g_s}(RM_*)^{1+d/2}\,.
\ee
 We recall that, in the absence of the dilaton $(g_s=1)$, the species scale associated to the KK tower of $d$ flat extra dimensions is of the order of the $(4+d)$-dimensional Planck scale $M_*$ and the number of species is given by $N=(RM_*)^d$, so that \eqref{M*} is reduced to the well known relation $M_*=M_p/\sqrt{N}$ \cite{Dvali:2007hz, Dvali:2007wp}. On the other hand for a perturbative string tower alone $(RM_*=1)$, the species scale is the string scale (still denoted $M_*$) and the number of species is given by $N=1/g_s^2$ \cite{Dvali:2009ks, Dvali:2010vm}. In the presence of the dilaton with a (modestly) perturbative string coupling, the 
string scale is slightly below the higher-dimensional Planck scale by a factor of $g_s^{1/d}$.

As a result, the compactification scale associated to the tower of `light' KK states is $R^{-1}\sim g_s e^{-5}M_p$ which is about three orders of magnitude below the Planck mass, around $3 \times 10^{15}$ GeV. This is well above the scale of inflation (by two orders of magnitude) and does not have any effect in the effective field theory of inflation. Note however that a value of $\phi \simeq 10$ (corresponding to $T_R\simeq 10^4$) would bring the compactification scale near the inflation scale ($H\sim M/\sqrt{12}\sim 5\times 10^{-6} M_P$) and larger values of $\phi$ would invalidate the effective field theory. We would like to emphasize that starting with the Planck determination of $n_s$ and noting that in this class of models, $n_s \approx 1-2/N_*$, where 
\beq
N_*\simeq - \int^{\phi_*}_{\phi_{\rm end}} \frac{1}{\sqrt{2\epsilon}} d\phi
\label{Ne}
\eeq
a value of $n_s \approx 0.965$ implies, a value of $N_* \approx 55$, which in turn implies a value of $\phi_* \approx 5$, the exact number depending on the specific potential which enters the integration through $\epsilon(\phi)$. Therefore this class of models, is capable of matching the observational constraints, while avoiding the swampland.  

Note that it has already been pointed out in the literature that modest super-Planckian displacements are not problematic for the validity of the effective field theory, as long as the species scale remains above the Hubble scale~\cite{Scalisi:2018eaz, vandeHeisteeg:2023uxj}. Here we impose a stronger condition that the scale of the KK tower (compactification scale) should remain above the Hubble scale during inflation.

A similar argument can be made for the supersymmetric $R^2$ theory with K\"ahler potential \eqref{Kdil} and superpotential \eqref{susy1}), or equivalently to \eqref{charge2} in the $y,z$ field basis. At $C=0$, the K\"ahler potential \eqref{Kdil} coincides with the one obtained by compactification of the ten-dimensional supergravity theory on a six-dimensional manifold of volume ${\cal V}=\prod_1^3(T_i+\bar{T}_i)^{1/2}$ where $T_i$ are the complex moduli of three mutually orthogonal 4-cycles. In this case~\cite{Becker:2002nn}
\be
K=-2\ln{\cal V}=-\sum_i\ln(T_i+\bar{T}_i)=-3\ln(T+\bar{T})\,,
\ee 
where the last equality is valid when all the three 4-cycles have the same size. One therefore obtains the kinetic terms:
\be
\label{kinTstaro}
-3{\partial T\partial\bar{T}\over (T+\bar{T})^2}=-12\left({\partial R\over R}\right)^2-{3\over 4}{(\partial b)^2\over R^4}\,,
\ee
where now $2 T=R^4+ib$ in string units. Agreement between \eqref{kinTstaro} with \eqref{dimredd} now implies that $d=6$, consistent with the argument above. It follows that the canonically normalized inflaton field is now $\phi=2\sqrt{6}\ln R$, while the compactification scale of six extra dimensions is found using \eqref{kinTstaro} for $d=6$ to be $R^{-1}\sim g_s e^{-5\sqrt{2/3}}M_p$ which is less than three orders of magnitude below the Planck mass, around $8 \times 10^{15}$ GeV. In this case, the effective field theory of inflation breaks around $\phi \sim 12$ again corresponding to $T_R\simeq 10^4$.

We conclude this section by deriving a relation between the number of e-folds $N_e$ during inflation and the number of species $N$, as advocated in the introduction. $N_e$ is directly related to the initial value of the inflaton through Eq.~(\ref{Ne}) by replacing $\phi_\star$ with the initial inflaton field value at the beginning of inflation, $\phi_0$, and by replacing $N_\star$ with $N_e$. More precisely, in the examples studied above with scalar potential \eqref{starpot} in the Starobinsky model and \eqref{ANRpot} for the string model, on obtain (assuming $\phi_{\rm end} \approx 0)$:
\beq\label{Nephi0}
N_e(\phi_0) \simeq 
\begin{cases}
\frac34 \left( e^{\sqrt{2/3} \phi_0} - \sqrt{\frac23} \phi_0 \right) \simeq \frac34  e^{\sqrt{2/3} \phi_0} \qquad {\rm from~Eq.~\eqref{starpot}} \\
\frac12  \left( e^{ \phi_0} -  \phi_0 \right) \simeq \frac12  e^{ \phi_0}  \qquad {\rm from~Eq.~\eqref{ANRpot}}
\end{cases}
\eeq
On the other hand $\phi_0$ determines the number of species $N$ as follows (assuming for simplicity $\phi_\text{end}\approx 0$).
Since $\phi_0>\phi_\star$ is large in Planck units, the KK tower opens up at a lower scale than estimated above, given by \eqref{RMp} with $\phi_\star$ replaced by $\phi_0$. The species scale is of the order of the higher dimensional Planck scale $M_*$ and the number of species is given by $N=(RM_*)^d$, where $d$ is the number of extra dimensions in the KK tower. From Eq.~(50) one obtains that $\phi=[d(d+2)/2]\ln (RM_*)$ and thus 
\beq\label{Nphi0}
N=e^{2\phi_0/(d+2)}.
\eeq
Combining \eqref{Nephi0} with \eqref{Nphi0}, one obtains a relation between $N_e$ and $N$, of the form $N_e\sim N^b$ where the exponent $b$ is of order unity and $b\gtrsim 1$, at least for the models considered here.

\section{Conclusions}
\label{sec:6}

Inflation has moved from being a paradigm to a testable theory.
Predictions for the space-time curvature and tilt of the CMB anisotropy
spectrum have been tested, and agree remarkably well with one of the 
first models of inflation, namely the Starobinsky model \cite{Starobinsky:1980te} based on 
an extension of Einstein gravity, to one where the gravitational 
Lagrangian contains a term quadratic in curvature. On the horizon,
is a test (or discovery) of the tensor-to-scalar ratio of primordial fluctuations,
predicted to be roughly an order of magnitude below current experimental limits.

It is well established that the Starobinsky model can be constructed
within the framework of no-scale supergravity \cite{Cecotti:1987sa,eno6}. Embedding the theory in
the context of string theory represents a bigger challenge. For example,
the identity of the inflaton (or scalaron in the $R+R^2$ theory),  and the scale of inflation are challenging questions.
In addition, since these constructions inevitably require additional scalar fields,
the theory must be stable so that the inflationary trajectory (in field space) 
follows that of the Starobinsky model. The model also necessarily involves large excursions
in field space which has been called into question as to whether such a theory
can be derived from string theory or is relegated to the swampland \cite{Ooguri:2006in}. Finally,
what is the role of the dilaton in Starobinsky inflation?
We have attempted to answer these questions in this work. 

We have seen in fact, that the incorporation of the dilaton in the $R + \alpha R^2$ theory is
actually rather straight forward. The parameter $\alpha$ can simply be made field dependent, 
with $\alpha \to \alpha e^{-2\varphi} = \alpha S$. The scale of inflation must be determined
in a specific model and here we discussed the free-fermionic construction of \cite{Antoniadis:2020txn},
where the scale of inflation is determined at 6th order in the perturbative expansion in $\alpha'$,
so that $M \sim 10^{-5} M_P$. This theory is actually Starobinsky like
(similar to an $\alpha$-Starobinsky model of~\cite{Ellis:2020xmk,Kallosh:2014rga,ENOV3} with $\alpha = 2/3$). 
The identity of the inflaton is now associated with the area modulus of a 2-cycle, $T$ in 
the string model. 

As noted several times, in addition to a (complex)-inflaton, an additional complex 
scalar is required. This can be seen from the supersymmetrization of the $R + \alpha R^2$
theory \cite{Cecotti:1987sa}, or from the requirement of achieving inflation within the framework of no-scale supergravity \cite{Ellis:2013nxa}. 
All through inflation, the additional directions must sit in stable minima. 
If the inflaton is associated with the real part of $T$, $\TR$, then $\TI = 0$ is automatically fixed. 
However, the additional field, $C$, (associated with the sgoldstino) typically has
$m_C^2 < 0$ and is in fact unstable. We have derived here general conditions
for which the dilaton can be used to stabilize the sgoldstino without 
any additional $C$-dependent corrections to either the K\"ahler potential or the superpotential.
We have also shown that in this construction, the necessary field excursions, $\phi \sim 5$
and $T\sim 75$ are small enough so as to evade the problems which relegate
large field inflation models to the swampland, thus turning the swampland into a mirage.

Before  
ending this paper, we comment on the necessity of fixing the VEV and stabilizing the dilaton.
In all of the arguments made above regarding the stabilization of the sgoldstino $C$, we had implicitly assumed that the VEV of $S$ was held fixed. However stabilizing the dilaton 
is not a new problem \cite{Damour:1994zq,Binetruy:1996xja,Dvali:1997ke,Abel:2000tf,Skinner:2003bi,Higaki:2003jt}. Indeed, it was argued \cite{Kalara:1988yu} that from the equations of motion, de Sitter-like solutions and inflation require dilaton self-interactions, and a stabilization mechanism.
It is beyond the scope of this work to resolve this long
standing problem. However, we would like to make a few observations. 
Adding a function, $g(S)$ to the superpotential does not work because although the mass-squared of $C$
is positive at the minimum when $\TR = 1/2$, generally at some large value of $\TR$ (i.e., during inflation), it turns negative upsetting the inflationary trajectory.
A possible way out is to separate the stabilization procedure in two steps by analogy with type IIB flux compactifications, where the dilaton and complex structure moduli are stabilized in a supersymmetric way prior to the K\"ahler class moduli that provide the inflaton potential~\cite{Kachru:2003aw,Kallosh:2004yh}.
Another possibility would be if some dynamics or supersymmetry breaking in the dilaton sector could provide a soft mass term for the canonically normalized dilaton $\Delta V = \frac12 M'^2 |\ln S/S_0|^2$, then all fields remain stabilized throughout the duration of inflation.  So long as $M' > M$, the dilaton remains fixed very near $S_0$. We hope to be able to return to this question in future work.

This string-derived realization of inflation, like the Starobinsky model, makes a definite prediction for the
tensor-to-scalar ratio, $r \simeq 0.0024$. This is slightly lower than the
standard Starobinsky model but still testable in the foreseeable future with
the next generation of CMB experiments. 

\vspace{-0.5cm}
\section*{Acknowledgements}
The work of I.A. is supported by the Second Century Fund (C2F), Chulalongkorn University. The work of K.A.O. was supported in part by DOE grant DE-SC0011842 at the University of Minnesota.


\begin{thebibliography}{99}

\bibitem{Starobinsky:1980te}
A.~A.~Starobinsky,
``A New Type of Isotropic Cosmological Models Without Singularity,''
Phys. Lett. B \textbf{91} (1980), 99-102.

\bibitem{Planck}
  N.~Aghanim \textit{et al.} [Planck],
``Planck 2018 results. VI. Cosmological parameters,''
Astron. Astrophys. \textbf{641}, A6 (2020)
[arXiv:1807.06209 [astro-ph.CO]].

\bibitem{Stelle:1976gc}
K.~S.~Stelle,
``Renormalization of Higher Derivative Quantum Gravity,''
Phys. Rev. D \textbf{16} (1977), 953-969

\bibitem{Whitt:1984pd}
B.~Whitt,
``Fourth Order Gravity as General Relativity Plus Matter,''
Phys. Lett. B \textbf{145} (1984), 176-178.

\bibitem{Ooguri:2006in}
H.~Ooguri and C.~Vafa,
``On the Geometry of the String Landscape and the Swampland,"
Nucl. Phys. B \textbf{766}, 21-33 (2007)
[arXiv:hep-th/0605264 [hep-th]].

\bibitem{Lee:2019wij}
S.~J.~Lee, W.~Lerche and T.~Weigand,
``Emergent strings from infinite distance limits,''
JHEP \textbf{02} (2022), 190
[arXiv:1910.01135 [hep-th]].

\bibitem{Higuchi:1986py}
A.~Higuchi,
``Forbidden mass range for spin-2 field theory in de Sitter space-time,"
Nucl. Phys. B \textbf{282} (1987), 397-436.

\bibitem{Cecotti:1987sa}
S.~Cecotti,
``Higher dervivative supergravity is equivalent to standard supergravity coupled to matter. 1.,''
Phys. Lett. B \textbf{190} (1987), 86-92

\bibitem{eno6}
   J.~Ellis, D.~V.~Nanopoulos and K.~A.~Olive,
  ``No-Scale Supergravity Realization of the Starobinsky Model of Inflation,''
  Phys.\ Rev.\ Lett.\  {\bf 111} (2013) 111301 
  [arXiv:1305.1247 [hep-th]].


  \bibitem{Ellis:2013nxa}
J.~Ellis, D.~V.~Nanopoulos and K.~A.~Olive,
``Starobinsky-like Inflationary Models as Avatars of No-Scale Supergravity,''
JCAP \textbf{10} (2013), 009
[arXiv:1307.3537 [hep-th]].

\bibitem{kl}
R.~Kallosh and A.~Linde,
  ``Superconformal generalizations of the Starobinsky model,''
  JCAP {\bf 1306} (2013) 028
  [arXiv:1306.3214 [hep-th]];\\
  ``Non-minimal Inflationary Attractors,''
  JCAP {\bf 1310} (2013) 033
  [arXiv:1307.7938].

\bibitem{Ellis:2020lnc}
For a review of no-scale supergravity realizations, see
J.~Ellis, M.~A.~G.~Garcia, N.~Nagata, D.~V.~Nanopoulos, K.~A.~Olive and S.~Verner,
``Building models of inflation in no-scale supergravity,''
Int. J. Mod. Phys. D \textbf{29}, no.16, 2030011 (2020)
[arXiv:2009.01709 [hep-ph]].



\bibitem{Antoniadis:2020txn}
I.~Antoniadis, D.~V.~Nanopoulos and J.~Rizos,
``Cosmology of the string derived flipped $SU(5)$,''
JCAP \textbf{03} (2021), 017
[arXiv:2011.09396 [hep-th]].

\bibitem{Antoniadis:1986rn}
  I.~Antoniadis, C.~P.~Bachas and C.~Kounnas,
  ``Four-Dimensional Superstrings,''
  Nucl.\ Phys.\ B {\bf 289} (1987) 87.
 
  \bibitem{Antoniadis:1987wp}
  I.~Antoniadis and C.~Bachas,
  ``4-D Fermionic Superstrings with Arbitrary Twists,''
  Nucl.\ Phys.\ B {\bf 298} (1988) 586.

  \bibitem{Barr}
S.~M.~Barr,
  Phys.\ Lett.\  {\bf 112B} (1982) 219;
  S.~M.~Barr,
  Phys.\ Rev.\ D {\bf 40}, 2457 (1989).

\bibitem{DKN}
J.~P.~Derendinger, J.~E.~Kim and D.~V.~Nanopoulos,
  Phys.\ Lett.\  {\bf 139B} (1984) 170.


\bibitem{Antoniadis:1989zy}
I.~Antoniadis, J.~R.~Ellis, J.~S.~Hagelin and D.~V.~Nanopoulos,
``The Flipped SU(5) x U(1) String Model Revamped,''
Phys. Lett. B \textbf{231} (1989), 65-74

\bibitem{Ellis:2018moe}
J.~Ellis, M.~A.~G.~Garcia, N.~Nagata, D.~V.~Nanopoulos and K.~A.~Olive,
``Starobinsky-like Inflation, Supercosmology and Neutrino Masses in No-Scale Flipped SU(5),''
JCAP \textbf{07} (2017) 006
[arXiv:1704.07331 [hep-ph]];\\
  ``Symmetry Breaking and Reheating after Inflation in No-Scale Flipped SU(5),''
  JCAP {\bf 1904} (2019) 009
  [arXiv:1812.08184 [hep-ph]];\\
  ``Cosmology with a master coupling in flipped SU(5) $\times$ U(1): the $\lambda_6$ universe,''
  Phys.\ Lett.\ B {\bf 797} (2019) 134864
  [arXiv:1906.08483 [hep-ph]];\\
``Superstring-Inspired Particle Cosmology: Inflation, Neutrino Masses, Leptogenesis, Dark Matter \& the SUSY Scale,''
  JCAP {\bf 2001} (2020) 035
  [arXiv:1910.11755 [hep-ph]].

     \bibitem{ENO8}
  J.~Ellis, D.~V.~Nanopoulos and K.~A.~Olive,
  ``A no-scale supergravity framework for sub-Planckian physics,''
  Phys.\ Rev.\ D {\bf 89} (2014) 4,  043502
  [arXiv:1310.4770 [hep-ph]];

\bibitem{Ellis:2018zya}
  J.~Ellis, D.~V.~Nanopoulos, K.~A.~Olive and S.~Verner,
  ``A general classification of Starobinsky-like inflationary avatars of SU(2,1)/SU(2) $\times$ U(1) no-scale supergravity,''
  JHEP {\bf 1903} (2019) 099
  [arXiv:1812.02192 [hep-th]].

\bibitem{Lopez:1994ej}
J.~L.~Lopez, D.~V.~Nanopoulos and K.~J.~Yuan,
``Moduli and Kahler potential in fermionic strings,''
Phys. Rev. D \textbf{50} (1994), 4060-4074
[arXiv:hep-th/9405120 [hep-th]].

\bibitem{Antoniadis:1987zk}
  I.~Antoniadis, J.~R.~Ellis, E.~Floratos, D.~V.~Nanopoulos and T.~Tomaras,
  ``The Low-energy Effective Field Theory From Four-dimensional Superstrings,''
  Phys.\ Lett.\ B {\bf 191} (1987) 96.


  




\bibitem{Ferrara:1987tp}
  S.~Ferrara, L.~Girardello, C.~Kounnas and M.~Porrati,
  ``Effective Lagrangians for Four-dimensional Superstrings,''
  Phys.\ Lett.\ B {\bf 192} (1987) 368;\\
  ``The Effective Interactions of Chiral Families in Four-dimensional Superstrings,''
  Phys.\ Lett.\ B {\bf 194} (1987) 358.

\bibitem{Dine:1987xk}
  M.~Dine, N.~Seiberg and E.~Witten,
  ``Fayet-Iliopoulos Terms in String Theory,''
  Nucl.\ Phys.\ B {\bf 289} (1987) 589.

\bibitem{Lopez:1989fb}
J.~L.~Lopez and D.~V.~Nanopoulos,
``Hierarchical Fermion Masses and Mixing Angles From the Flipped String,''
Nucl. Phys. B \textbf{338} (1990), 73-100;\\
``Decisive role of nonrenormalizable terms in the flipped string,''
Phys. Lett. B \textbf{251} (1990), 73-82;\\
``Sharpening the flipped SU(5) string model,''
Phys. Lett. B \textbf{268} (1991), 359-364.

\bibitem{Rizos:1990xn}
J.~Rizos and K.~Tamvakis,
``Retracing the phenomenology of the flipped SU(5) x U(1) superstring model,''
Phys. Lett. B \textbf{251} (1990), 369-378.

\bibitem{Ellis:1990vy}
J.~R.~Ellis, J.~L.~Lopez and D.~V.~Nanopoulos,
``Baryon decay: Flipped SU(5) surmounts another challenge,''
Phys. Lett. B \textbf{252} (1990), 53-58.

\bibitem{Kalara:1991qh}
S.~Kalara, J.~L.~Lopez and D.~V.~Nanopoulos,
``Gauge and matter condensates in realistic string models,''
Phys. Lett. B \textbf{275} (1992), 304-310
[arXiv:hep-th/9110023 [hep-th]].

\bibitem{Antoniadis:1991fc}
I.~Antoniadis, J.~Rizos and K.~Tamvakis,
``Gauge symmetry breaking in the hidden sector of the flipped SU(5) x U(1) superstring model,''
Phys. Lett. B \textbf{278} (1992), 257-265;\\
``Naturally light neutrinos in the flipped SU(5) x U(1) superstring model,''
Phys. Lett. B \textbf{279} (1992), 281-284.

\bibitem{Antoniadis:2021rfm}
I.~Antoniadis, D.~V.~Nanopoulos and J.~Rizos,
``Particle physics and cosmology of the string derived no-scale flipped SU(5),''
Eur. Phys. J. C \textbf{82} (2022) no.4, 377
[arXiv:2112.01211 [hep-th]].


\bibitem{Ferrara:2013wka}
S.~Ferrara, R.~Kallosh and A.~Van Proeyen,
``On the Supersymmetric Completion of $R+R^2$ Gravity and Cosmology,''
JHEP \textbf{11} (2013), 134
[arXiv:1309.4052 [hep-th]].

\bibitem{Antoniadis:2014oya}
I.~Antoniadis, E.~Dudas, S.~Ferrara and A.~Sagnotti,
``The Volkov\textendash{}Akulov\textendash{}Starobinsky supergravity,''
Phys. Lett. B \textbf{733} (2014), 32-35
[arXiv:1403.3269 [hep-th]].

\bibitem{Casas:2024jbw}
G.~F.~Casas and L.~E.~Ib\'a\~nez,
``Modular Invariant Starobinsky Inflation and the Species Scale,''
[arXiv:2407.12081 [hep-th]].




\bibitem{Rasulian:2021wny}
I.~M.~Rasulian, M.~Torabian and L.~Velasco-Sevilla,
``Swampland de Sitter conjectures in no-scale supergravity models,''
Phys. Rev. D \textbf{104}, no.4, 044028 (2021)
[arXiv:2105.14501 [hep-th]].

\bibitem{Scalisi:2018eaz}
M.~Scalisi and I.~Valenzuela,
``Swampland distance conjecture, inflation and $\alpha$-attractors,''
JHEP \textbf{08} (2019), 160
[arXiv:1812.07558 [hep-th]].

\bibitem{Lust:2023zql}
D.~L\"ust, J.~Masias, B.~Muntz and M.~Scalisi,
``Starobinsky inflation in the swampland,''
JHEP \textbf{07} (2024), 186
[arXiv:2312.13210 [hep-th]].

\bibitem{Kallosh:2024ymt}
R.~Kallosh and A.~Linde,
``$SL(2,\mathbb{Z})$ Cosmological Attractors,''
[arXiv:2408.05203 [hep-th]].


\bibitem{wesbag}
J. Wess, J. Bagger, \textit{Supersymmetry and Supergravity,} Princeton University Press, Princeton, New Jersey, 1992.



\bibitem{Cremmer:1983bf}
  E.~Cremmer, S.~Ferrara, C.~Kounnas and D.~V.~Nanopoulos,
  ``Naturally Vanishing Cosmological Constant in $ \mathcal{N}=1 $ Supergravity,''
  Phys.\ Lett.\  {\bf 133B} (1983) 61.



\bibitem{Ellis:1983ei}
  J.~R.~Ellis, C.~Kounnas and D.~V.~Nanopoulos,
  ``Phenomenological SU(1,1) Supergravity,''
  Nucl.\ Phys.\ B {\bf 241} (1984) 406;\\
``No Scale Supersymmetric Guts,''
Nucl. Phys. B \textbf{247} (1984), 373-395.

\bibitem{Lahanas:1986uc}
for a review, see
A.~B.~Lahanas and D.~V.~Nanopoulos,
``The Road to No Scale Supergravity,''
Phys. Rept. \textbf{145} (1987), 1

 \bibitem{DLT}
G.~D.~Diamandis, A.~B.~Lahanas and K.~Tamvakis,
  Phys.\ Rev.\ D {\bf 92} (2015) no.10,  105023
  [arXiv:1509.01065 [hep-th]].

  \bibitem{DGKLP}
   G.~A.~Diamandis, B.~C.~Georgalas, K.~Kaskavelis, A.~B.~Lahanas and G.~Pavlopoulos,
  Phys.\ Rev.\ D {\bf 96}, no. 4, 044033 (2017)
  [arXiv:1704.07617 [hep-th]].
 

  
  
\bibitem{eno9}
J.~Ellis, D.~V.~Nanopoulos and K.~A.~Olive,
Phys. Rev. D \textbf{97}, no.4, 043530 (2018)
[arXiv:1711.11051 [hep-th]].

\bibitem{Ema:2024sit}
Y.~Ema, M.~A.~G.~Garcia, W.~Ke, K.~A.~Olive and S.~Verner,
Universe \textbf{10}, no.6, 239 (2024)
[arXiv:2404.14545 [hep-ph]].

\bibitem{Ellis:1984bs}
J.~R.~Ellis, C.~Kounnas and D.~V.~Nanopoulos,
``No Scale Supergravity Models with a Planck Mass Gravitino,''
Phys. Lett. B \textbf{143}, 410-414 (1984)

\bibitem{Ellis:2015kqa}
J.~Ellis, M.~A.~G.~Garcia, D.~V.~Nanopoulos and K.~A.~Olive,
``Phenomenological Aspects of No-Scale Inflation Models,''
JCAP \textbf{10}, 003 (2015)
[arXiv:1503.08867 [hep-ph]].

\bibitem{Kaneta:2019yjn}
K.~Kaneta, Y.~Mambrini, K.~A.~Olive and S.~Verner,
``Inflation and Leptogenesis in High-Scale Supersymmetry,''
Phys. Rev. D \textbf{101}, no.1, 015002 (2020)
[arXiv:1911.02463 [hep-ph]].

\bibitem{Ellis:2020xmk}
J.~Ellis, D.~V.~Nanopoulos, K.~A.~Olive and S.~Verner,
``Phenomenology and Cosmology of No-Scale Attractor Models of Inflation,''
JCAP \textbf{08}, 037 (2020)
[arXiv:2004.00643 [hep-ph]].



\bibitem{KLR}
R.~Kallosh, A.~Linde and D.~Roest,
``Superconformal Inflationary $\alpha$-Attractors,''
JHEP \textbf{11}, 198 (2013)
[arXiv:1311.0472 [hep-th]].

\bibitem{Kallosh:2014rga}
R.~Kallosh, A.~Linde and D.~Roest,
``Large field inflation and double $\alpha$-attractors,''
JHEP \textbf{08}, 052 (2014)
[arXiv:1405.3646 [hep-th]].


\bibitem{Roest:2015qya}
D.~Roest and M.~Scalisi,
Phys. Rev. D \textbf{92}, 043525 (2015)
[arXiv:1503.07909 [hep-th]].

\bibitem{Scalisi:2015qga}
M.~Scalisi,
JHEP \textbf{12}, 134 (2015)
[arXiv:1506.01368 [hep-th]].



\bibitem{ENOV3}
J.~Ellis, D.~V.~Nanopoulos, K.~A.~Olive and S.~Verner,
``Unified No-Scale Attractors,''
JCAP \textbf{09}, 040 (2019)
[arXiv:1906.10176 [hep-th]].



\bibitem{Saini:2024mun}
S.~Saini and A.~Nautiyal,
``Observational constraints on $\alpha$-Starobinsky inflation,''
[arXiv:2409.05615 [astro-ph.CO]].



\bibitem{Dvali:2007hz}
G.~Dvali,
``Black holes and large $N$ species solution to the hierarchy problem,"
Fortsch. Phys. \textbf{58}, 528-536 (2010)
[arXiv:0706.2050 [hep-th]].
 
\bibitem{Dvali:2007wp}
G.~Dvali and M.~Redi,
``Black hole bound on the number of species and quantum gravity at LHC,"
Phys. Rev. D \textbf{77}, 045027 (2008)
[arXiv:0710.4344 [hep-th]].
 
 
 
\bibitem{Dvali:2009ks}
G.~Dvali and D.~L\"ust,
``Evaporation of Microscopic Black Holes in String Theory and the Bound on Species,"
Fortsch. Phys. \textbf{58} (2010), 505-527
[arXiv:0912.3167 [hep-th]].
 
 
\bibitem{Dvali:2010vm}
G.~Dvali and C.~Gomez,
``Species and Strings,"
[arXiv:1004.3744 [hep-th]].

\bibitem{vandeHeisteeg:2023uxj}
 D.~van de Heisteeg, C.~Vafa, M.~Wiesner and D.~H.~Wu,
 ``Bounds on field range for slowly varying positive potentials,''
JHEP \textbf{02} (2024), 175
[arXiv:2305.07701 [hep-th]].

 
 	\bibitem{Becker:2002nn} See e.g.
	K.~Becker, M.~Becker, M.~Haack and J.~Louis,
	``Supersymmetry breaking and alpha-prime corrections to flux induced potentials,''
	JHEP {\bf 0206} (2002) 060
	[hep-th/0204254].

\bibitem{Damour:1994zq}
T.~Damour and A.~M.~Polyakov,
``The String dilaton and a least coupling principle,''
Nucl. Phys. B \textbf{423} (1994), 532-558
[arXiv:hep-th/9401069 [hep-th]].

\bibitem{Binetruy:1996xja}
P.~Binetruy, M.~K.~Gaillard and Y.~Y.~Wu,
``Dilaton stabilization in the context of dynamical supersymmetry breaking through gaugino condensation,''
Nucl. Phys. B \textbf{481}, 109-128 (1996)
[arXiv:hep-th/9605170 [hep-th]].

\bibitem{Dvali:1997ke}
G.~R.~Dvali and Z.~Kakushadze,
``A Remark on dilaton stabilization,''
Phys. Lett. B \textbf{417}, 50-52 (1998)
[arXiv:hep-th/9709093 [hep-th]].

\bibitem{Abel:2000tf}
S.~A.~Abel and G.~Servant,
``Dilaton stabilization in effective type I string models,''
Nucl. Phys. B \textbf{597}, 3-22 (2001)
[arXiv:hep-th/0009089 [hep-th]].

\bibitem{Skinner:2003bi}
D.~Skinner,
``Inflation and Kahler stabilization of the dilaton,''
Phys. Rev. D \textbf{67}, 103506 (2003)
[arXiv:hep-th/0302024 [hep-th]].


\bibitem{Higaki:2003jt}
T.~Higaki, Y.~Kawamura, T.~Kobayashi and H.~Nakano,
``Non-perturbative K\"ahler potential, dilaton stabilization and Fayet\textendash{}Iliopoulos term,''
Phys. Lett. B \textbf{582}, 257-262 (2004)
[arXiv:hep-ph/0311315 [hep-ph]].


\bibitem{Kalara:1988yu}
S.~Kalara and K.~A.~Olive,
``Difficulties for Field Theoretical Inflation in String Models,''
Phys. Lett. B \textbf{218}, 148 (1989).

\bibitem{Kachru:2003aw}
  S.~Kachru, R.~Kallosh, A.~D.~Linde and S.~P.~Trivedi,
``De Sitter vacua in string theory,''
  Phys.\ Rev.\  D {\bf 68}, 046005 (2003)
  [arXiv:hep-th/0301240].

\bibitem{Kallosh:2004yh}
  R.~Kallosh and A.~D.~Linde,
``Landscape, the scale of SUSY breaking, and inflation,''
  JHEP {\bf 0412}, 004 (2004)
  [arXiv:hep-th/0411011];\\
``Testing String Theory with CMB,''
  JCAP {\bf 0704}, 017 (2007)
  [arXiv:0704.0647 [hep-th]].





\end{thebibliography}
\end{document}